\documentclass{mem}
\usepackage{natbib}\usepackage{txfonts}\usepackage{balance}
\usepackage{graphicx}
\usepackage[a4paper]{hyperref}
\idline{75}{282}
\begin{document}
\def\teff{$T\rm_{eff }$}
\def\kms{$\mathrm {km s}^{-1}$}

\title{Variability and stability in optical blazar jets}
\subtitle{The case of OJ287}

\author{
Carolin Villforth\inst{1,2}
\and
Kari Nilsson\inst{2}
\and
Jochen Heidt\inst{3}
\and
Tapio Pursimo\inst{4}
}

\offprints{C. Villforth (carovi@utu.fi,villfort@stsci.edu)}

\institute{
Space Telescope Science Institute,
3700 San Martin Drive,
Baltimore, MD 21218, USA
\email{carovi@utu.fi,villfort@stsci.edu}
\and
University of Turku, Department of Physics and Astronomy, Tuorla Observatory,
V\"{a}is\"{a}l\"{a}ntie 20,
21500  Piikki\"{o},
Finland
\and
ZAH, Landessternwarte Heidelberg,
Kšnigstuhl 12,
69117 Heidelberg,
Germany
\and
Nordic Optical Telescope,
Apartado 474,
38700 Santa Cruz de La Palma,
Spain
}
\authorrunning{Villforth et al.}

\titlerunning{Variability and stability in optical blazar jets}

\abstract{OJ287 is a BL Lac object at redshift $z=0.306$ that has shown
double-peaked bursts at regular intervals of $\sim 12$ yr during the last $\sim
40$ yr. Due to this behavior, it has been suggested that OJ287 might host a close
supermassive binary black hole. We present optical photopolarimetric monitoring
data from 2005--2009, during which the latest double-peaked outburst occurred. We
find a stable component in the optical jet: the optical polarization core. The
optical polarization indicates that the magnetic field is oriented parallel to
the jet. Using historical optical polarization data, we trace the evolution of
the optical polarization core and find that it has showed a swing in the Stokes plane
indicating a reorientation of the jet magnetic field. We also find that changes
in the optical jet magnetic field seem tightly related to the double-peaked
bursts. We use our findings as a new constraint on possible binary black hole
models. Combining all available observations, we find that none of the proposed
binary black bole models is able to fully explain the observations. We suggest a
new approach to understanding OJ287 that is based on the assumption that changes in
the jet magnetic field drive the regular outbursts.
 }
\maketitle{}

\section{Introduction}

Blazars are amongst the most violently variable sources in the Universe.
According to the standard model \citep{urry_unified_1995}, blazars are AGN with a
jet pointing almost directly towards the observer, the jet radiation is thus
highly beamed and dominates the spectrum. Therefore, blazars are perfect
laboratories to study variability and turbulence in AGN jets.

Blazars show variability on time scales from hours to decades, with partially
extreme amplitudes
\citep{ulrich_variability_1997,valtaoja_radio_2000,villforth_intranight_2009}.
The radio structure of blazar jets consists of a so-called
radio-core that does not move and blobs that appear to be ejected from the core
and move away from it at apparently superluminal speeds
\citep[e.g.][]{jorstad_multiepoch_2001}. It is assumed that the core represents a
standing shock front at the end of the collimination zone
\citep{marscher_jets_2009}. The highest energy photons are believed to originate
close to the black hole, with x-ray and optical emission originating in the
collimination zone \citep{marscher_jets_2009}. Radio emission dominates further
downstream.

This picture agrees rather well withy magneto-hydrodynamical simulations
\citep{nakamura_production_2001}. Those simulations also indicate that the
magnetic field is ordered in the collimination zone, building a toroidal magnetic
field. At the end of the collimination zone, a standing shock front appears and
further down the jet the toroidal jet magnetic field breaks up due to
turbulences. This theoretical expectation seems to describe the observations in
the jet of the nearby radio galaxy M87 rather well
\citep{biretta_radio_1991,junor_formation_1999,asada_evn_2008}.

Given the fact that most of the studies on blazar jets have been performed in the
radio \citep[e.g.][]{jorstad_multiepoch_2001,gabuzda_nature_2003}, optical
polarization studies can give new insights into the parts of blazar jets in which
the optical emission originates
\citep[e.g.][]{hagen-thorn_oj_1980,holmes_polarization_1984,jannuzi_optical_1994,gabuzda_evidence_2006,villforth_variability_2010}.

For our study, we choose OJ287, which is one of the best studied blazars. OJ287
has received a lot of attention as it has shown massive double-peaked outbursts
approximately every 12 years during the last 40--100 years
\citep{villforth_variability_2010}. It has been therefore been suggested that
OJ287 hosts a close supermassive binary black hole
\citep[e.g.][]{sillanp_oj_1988,lehto_oj_1996,katz_precessing_1997,villata_beaming_1998,valtaoja_radio_2000,valtonen_predictingnext_2006,valtonen_new_2007,valtonen_tidally_2009}.
Our monitoring campaign covers the last double-peaked outburst observed in
OJ287. We can therefore also use the data to constrain the binary black hole
models proposed for OJ287.

\section{Observations and Data Reduction}

\begin{figure*}
\resizebox{\hsize}{!}{\includegraphics[width=16cm]{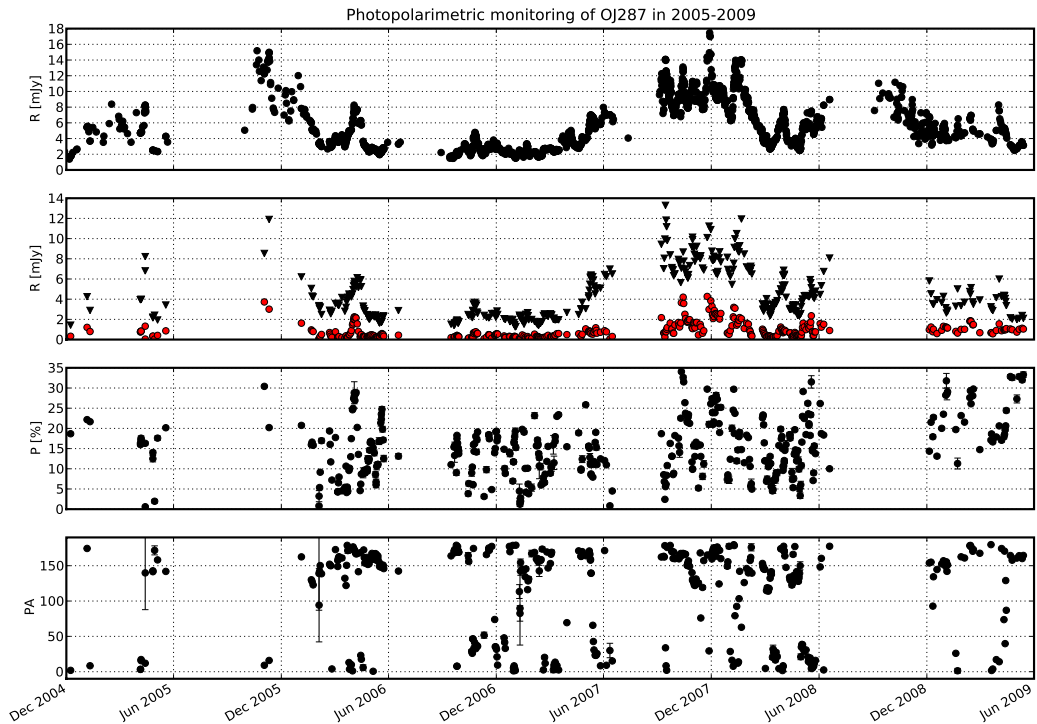}}
\caption{
Full photopolarimetric light-curve for OJ287 in 2005--2009. Panels show following
values, from top to bottom: total flux; polarized (red circles) and unpolarized
(black triangles) flux; degree of polarization; position angle.
}
\label{data}
\end{figure*}

This study is based on the data published in
\citep{villforth_variability_2010}. Information about observations and data
reduction can be found in said paper. All data are also available on Vizier
\citep{villforth_optical_2010}. A plot of the entire data is shown in Fig
\ref{data}.

\section{Using optical polarization data to trace the jet magnetic field}

As can be seen in Fig. \ref{data}, the position angle of the optical polarization
in OJ287 had a clear preferred value during our monitoring campaign. This is a
behavior that is often observed in BL Lac objects \citep{jannuzi_optical_1994}.
This behaviour is possibly less prominent in Flat Spectrum Radio Quasars (FSRQs)
\citep{angel_optical_1980,villforth_variability_2010}.

To study the origin of this preferred position angle, we study the
properties of the entire polarimetric data set in the Stokes plane. It can be
shown that there is a very clearly defined peak in the distributions of both Stokes
parameters. Assuming that the peak is caused by a single source of polarized
emission, the optical polarization core (OPC), we can subtract the OPC
vectorially from all data points. We find that the alignment in position angle
disappears. This means that the preferred position angle can be explained by a
single source of polarized emission, the OPC. Note that distributions in the
Stokes plane similar to the one observed in OJ287 are commonly observed in BL
Lac objects \citep{jannuzi_optical_1994}.

The emission of the OPC constitutes about 20\% of the optical emission of OJ287
in a moderate state, assuming a degree of polarization of 80\%. This is the
maximum degree of polarization possible in synchrotron radiation. We can
interpret this emission either as a sign of emission from a quiescent jet with a
global magnetic field or as a sign of a standing shock front. In
\cite{villforth_variability_2010} we argued that it is likely that the
emission  originates from a quiescent jet. This is also supported by the finding
that the degree of polarization and total optical flux show no correlation
\citep{villforth_variability_2010}. This finding is inconsistent with a model in
which all polarized emission originates in shock fronts
\citep{marscher_models_1985}. Therefore the quiescent jet hypothesis is favoured.


Given above findings, it is of interest to understand the long term evolution of
the optical polarization core.

\begin{figure*}
\resizebox{\hsize}{!}{\includegraphics[width=6cm]{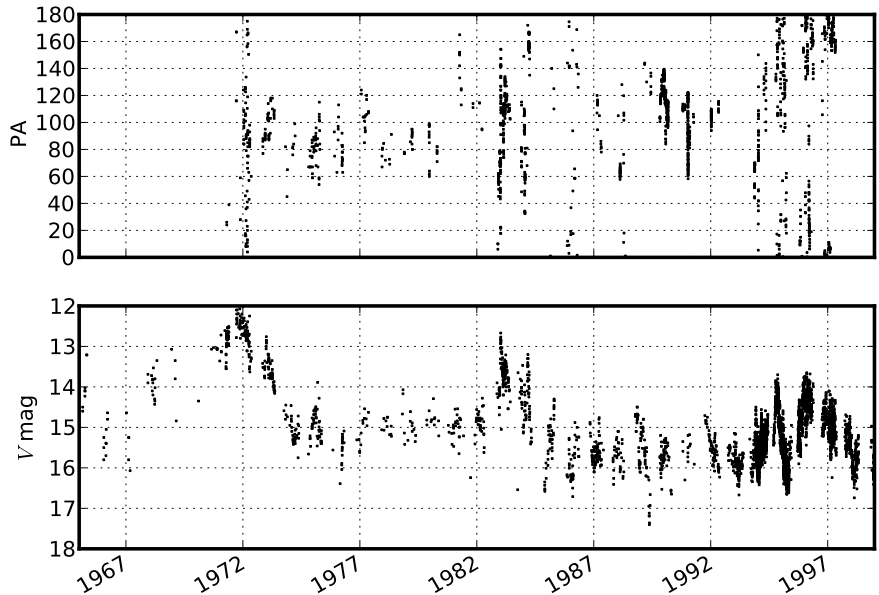}}
\caption{Evolution of the optical position angle $PA$ (upper panel) and and
optical magnitude (lower panel) of OJ287 from $\sim$ 1970 till the mid 1990s.}
\label{longPA}
\end{figure*}

We show a plot of the long term evolution of the optical polarization position
angle in Fig. \ref{longPA}. Note that this data is combined from data sets in
several filters. However, given that the differences between the position angle
in the optical are at most on the order of a few degrees
\citep{holmes_polarization_1984}, this should not affect our analysis.

We clearly see that the position angle changed its preferred direction, the
changes in the alignment of the optical polarization seem to be tightly related
to the outburst in optical flux.

\section{Does OJ287 host a supermassive binary black hole?}

Another open question is if OJ287 hosts a supermassive binary black hole. This
has been suggested by several authors
\citep[e.g.][]{sillanp_oj_1988,lehto_oj_1996,katz_precessing_1997,villata_beaming_1998,valtaoja_radio_2000,valtonen_predictingnext_2006,valtonen_tidally_2009}.
See \cite{villforth_variability_2010} for a summary and description of all these
models. Summarizing all available data, we find that none of the models is
fully able to explain all the observations.

Given that our findings of the OPC are indicating strong changes in the jet
magnetic field that are not consistent with any model based on a binary black
hole, we suggest a new approach for OJ287 models.

We suggest that the changes in the jet magnetic field are causing the double
peaked bursts. We suggest that resonance in the accretion disk magnetic field is
causing avalanche accretion of poloidal magnetic field lines, thereby changing
the pitch angel of the toroidal jet magnetic field. Over time, the change in the
pitch angle will change the apparent direction of the projected jet magnetic
field, as observed in OJ287 \citep{villforth_variability_2010}.

\section{Conclusions}

Based on about three years of photopolarimetric monitoring of the blazar OJ287,
we study the variability and stability of optical blazar jets. We also asses
different binary black hole models that have been proposed to explain the
regularly appearing double-peaked optical outbursts observed in OJ287.

We found a stable component in the optical polarization, the optical
polarization core (OPC). We interpret this as the emission from the quiescent
jet, the OPC can therefore be used  to trace the jet magnetic field.

Using all available data, we find that none of the proposed binary black hole
models is fully able to explain all the observations. Based on our finding of
strong changes in the jet magnetic field, we suggest that the double-peaked outbursts
observed in OJ287 are due to accretion of magnetic field, causing a change in
the pitch angle. This is observed as an apparent change in the projected jet
magnetic field.

Further observations will have to show if possible further outbursts can
constrain models on OJ287. Continued long term photopolarimetric monitoring will
be needed to achieve this goal. Till then, the case of OJ287 will remain a
mystery.

\begin{acknowledgements}
We would like to thank all observers for the OJ287 monitoring campaign.
\end{acknowledgements}

\bibliographystyle{aa}
\bibliography{Villforth}

\end{document}